\documentclass[12pt]{article}

\usepackage{amsmath,amssymb,amsfonts}
\usepackage{booktabs}
\usepackage{multirow}
\usepackage{array}
\usepackage{url}
\usepackage{graphicx}
\usepackage{xcolor}
\usepackage{float}
\usepackage{placeins}
\usepackage{geometry}
\usepackage{hyperref}
\usepackage{adjustbox}

\newcommand{\Gk}{G_k}
\newcommand{\Z}{\mathbb{Z}}
\newcommand{\dist}{\operatorname{dist}}
\newcommand{\modG}{\operatorname{mod}_{G_k}}

\newcommand{\Comp}{\mathcal{C}}
\newcommand{\Cgraph}{\mathcal{H}}

\newcommand{\Ball}{B_k}

\newcommand{\Fe}{F_E}
\newcommand{\Fv}{F_V}

\title{Re-Rooting-Assisted Edge-Minimum Runtime Repair for Node and Link Failures in Dense Gaussian Broadcast Networks}

\author{Bader A. Albader\\
\small Department of Computer Science, Faculty of Science, Kuwait University, Kuwait\\
\small \texttt{albader@cs.ku.edu.kw}}

\date{}

\begin{document}
\maketitle

\begin{abstract}
Dense Gaussian networks are degree-four algebraic interconnection networks with compact diameter and coordinate-based routing. Their diameter-level broadcast trees are efficient but fragile under node faults, link faults, and faults discovered during propagation. This paper develops a self-contained runtime recovery framework for dense Gaussian broadcast networks under static node/link faults, mixed static faults, and runtime-discovered single-link faults. The method first re-roots the effective source so known node faults become boundary leaves whenever possible, then filters failed links and repairs residual fragmentation by connecting the healthy components of the pruned tree. For a selected root with connected healthy component graph, we prove that exactly $c-1$ external component-crossing repair edges are necessary and sufficient. We also prove deterministic single-link repair, give a constant-size shifted boundary-intersection primitive for two-node source selection, derive a link-avoidance exclusion test, and add a local-obstruction probability bound explaining why fixed-size high-order cuts rapidly vanish as $k$ grows. Experiments over $k\in\{10,25,50,100,200\}$, up to 80,401 nodes, 280,000 static trials, and 15,000 transient trials show 100\% recovery for all deterministic and bounded mixed regimes, 99.998\% recovery for multi-link faults, and 99.963\% recovery for higher-order heuristic regimes; every non-recovered static trial is explained by disconnected component graphs or relocation failure. Compared with fixed-source component repair, re-rooting reduces average external repair edges by 80--100\%. Patched native Gaussian-link Noxim scheduled replays at $k=25$ and audited $k=50$ confirm packet-complete router-level execution and show that hybrid re-rooting sharply reduces repair edges, components, and repaired depth. A completion-cycle audit separates structural repair benefit from latency: zero-setup and latency-weighted ablations confirm that completion time depends on relocation/setup, replay scheduling, delivery tail, and selector objective, so the paper claims edge-minimum repair and depth reduction rather than universal completion-cycle dominance.
\end{abstract}


\section{Introduction}
One-to-all broadcast is a core collective operation in parallel processors, many-core systems, NoC fabrics, and distributed accelerators. It is used for synchronization, coherence messages, configuration updates, and control dissemination. A broadcast primitive in a low-degree topology should have small depth, avoid duplicate traffic, use simple forwarding state, and tolerate faults without global rebuilding after every small failure.

Dense Gaussian networks provide an algebraic topology in which this problem can be studied with explicit coordinate structure. For parameter $k$, the dense Gaussian network generated by $k+(k+1)i$ has
\begin{equation}
N=k^2+(k+1)^2=2k^2+2k+1
\end{equation}
nodes, degree four, and diameter $k$. Its canonical coordinate model identifies the vertices with the Manhattan ball
\begin{equation}
\Ball=\{(x,y)\in\Z^2: |x|+|y|\le k\}
\end{equation}
through the integer label
\begin{equation}
\phi(x+yi)\equiv kx+(k+1)y \pmod N.
\end{equation}
This model supports deterministic source-centered broadcast trees of depth at most $k$.

Prior static re-rooting work shows that one and two failed nodes can be neutralized by relocating the source to a node at distance $k$ from each failed node. Such faults become leaves of the broadcast tree and do not forward packets. Prior component-repair work shows that if a fault-pruned tree breaks into $c$ healthy components and the component graph is connected, then $c-1$ component-crossing repair edges are both necessary and sufficient. These results are useful, but neither alone solves runtime node/link failures. Re-rooting by itself does not repair residual failed tree links or high-order fault cuts. Component repair by itself can succeed from a fixed source but may require many repair edges if the fixed source leaves faults on major trunks.

This paper addresses the missing problem: \emph{runtime node/link broadcast recovery by re-rooting first and repairing only the residual damage}. The intellectual point is not a mechanical juxtaposition of two earlier techniques. The paper develops a new link-failure exclusion condition, a runtime micro-re-rooting protocol, a candidate-generation/selection interface, a root-cause analysis for non-perfect high-order recovery, and a quantitative repair-edge reduction study against fixed-source component repair. The combination changes the optimization objective: the selected root is not merely a valid source, but a damage-minimizing source that reduces the number of healthy components before the edge-minimum repair phase is invoked.

\subsection{Contributions}
The contributions are as follows.
\begin{itemize}
\item We formulate joint node/link fault-tolerant broadcasting in dense Gaussian networks as re-rooting-assisted component repair, covering static node/link faults, mixed faults, and runtime-discovered single-link faults.
\item We give a self-contained re-rooting primitive with the one/two node guarantee and a constant-size $9\times16$ shifted sign-case candidate generator for two-node source selection.
\item We introduce a link-failure exclusion test and prove an edge-minimum component-repair theorem: if the healthy component graph is connected after pruning, exactly $c-1$ component-crossing repair edges are necessary and sufficient.
\item We prove deterministic single-link repair and a high-order threshold result: one failed link is always avoided or repaired by one crossing edge, while three-node zero-repair re-rooting is governed by a triple boundary-intersection condition.
\item We add deterministic component-count and repair-edge bounds, distinguish repair state from repaired hop depth, and provide a local-obstruction probability bound explaining why fixed-size high-order cuts become rare as $k$ increases.
\item We evaluate 280,000 static trials and 15,000 transient trials up to 80,401 nodes, reporting success rates, candidate overhead, repair-edge reduction, transient discovery layer, and a complete root-cause breakdown of all non-recovered cases.
\item We validate the NoC interpretation using documented patched native Gaussian-link Noxim scheduled replays, a $k=25$ completion-cycle audit with zero-setup and latency-objective ablations, and a $k=50$ full-root-scan sanity check.
\end{itemize}

\subsection{Scope and Relationship to Related Manuscripts}
The present paper is self-contained: every theorem, selection rule, and repair procedure used in the experiments is stated here. Static node-failure re-rooting and multi-orientation component repair are discussed as related foundations, but the present contribution is the link/runtime hybrid framework and its analysis. In particular, the present paper studies failed links, runtime discovery, link-safety filtering, repair-edge reduction after re-rooting, and root-cause analysis of high-order failures; these are not consequences of either static re-rooting alone or fixed-source component repair alone. The companion-style references are therefore contextual rather than logical prerequisites for the correctness claims made here.

Table~\ref{tab:novelty_separation} makes this separation explicit. The present paper's new unit of analysis is the pair
\begin{equation}
(r,\Cgraph_r),
\end{equation}
namely the selected root together with the component graph induced after node and link pruning. The optimization is therefore not only to find a valid root or to repair a fixed tree, but to choose a root that makes the residual component graph small and repairable.

\begin{table}[H]
\centering
\caption{Novelty separation from related companion-style ingredients.}
\label{tab:novelty_separation}
\footnotesize
\begin{adjustbox}{max width=\textwidth}
\begin{tabular}{p{0.25\linewidth}p{0.31\linewidth}p{0.35\linewidth}}
\toprule
Ingredient & Solves & Not solved by that ingredient alone \\
\midrule
Static re-rooting & Places one/two failed nodes on the boundary. & Failed links, transient discovery, residual component cuts, repair-edge cost. \\
Component repair & Reconnects a chosen pruned tree with $c-1$ edges when $\Cgraph$ is connected. & How to choose a root that minimizes damage before repair. \\
This paper & Selects a damage-reducing root, filters link failures, repairs residual components, and explains non-perfect high-order cases. & Arbitrary clustered high-order faults without the connected-component condition. \\
\bottomrule
\end{tabular}
\end{adjustbox}
\end{table}

\section{Related Work}
\label{sec:related}
Algebraic interconnection networks use group or ring structure to obtain symmetry, small routing tables, and concise path descriptions. Akers and Krishnamurthy introduced a group-theoretic view of symmetric interconnection networks~\cite{Akers1989}. Classical texts describe meshes, tori, hypercubes, Cayley graphs, collective communication, deadlock-free routing, and fault tolerance in interconnection networks~\cite{Leighton1992,DallyTowles2004,Duato2003,Grama2003}. Dense Gaussian networks belong to this tradition: they are degree-four quotient networks over Gaussian integers with small diameter and coordinate-based routing.

Gaussian and circulant-based topologies have been studied for routing, resource placement, Hamiltonicity, and NoC suitability. Martinez et al. studied dense Gaussian networks as low-degree on-chip multiprocessor topologies~\cite{Martinez2006} and modeled toroidal networks using Gaussian integers~\cite{Martinez2008}. Flahive and Bose studied Gaussian and Eisenstein--Jacobi interconnection networks~\cite{FlahiveBose2010}. Additional work considered routing and path structures in dense Gaussian and optimal degree-four circulant networks~\cite{Alsaleh2013,Monakhova2023,Romanov2020}. Edge-disjoint Hamiltonian cycles in Gaussian networks provide another form of path diversity~\cite{AlbaderBose2016}.

Fault-tolerant broadcasting and reliable communication have often been addressed using independent spanning trees, completely independent spanning trees, redundant paths, or adaptive routing. Independent spanning trees provide internally disjoint delivery structures and are widely used as a protection mechanism in vertex- or edge-fault settings~\cite{ItaiRodeh1988,Hasunuma2002,Cheng2021}. Related tree-diversity approaches are attractive when steady-state redundancy is acceptable: the system maintains multiple delivery structures and switches traffic after a fault. The present paper targets a different point in the design space. It keeps one deterministic delivery tree in the fault-free case and pays only for the exceptional component-crossing rules needed after faults are observed. In NoC systems, fault tolerance has been studied using adaptive routing, spare links, region avoidance, reconfiguration, and application mapping~\cite{Kliazovich2013,PasrichaDutt2008,FlichBertozzi2010,Ebrahimi2013}. Such methods typically optimize packet routing, traffic distribution, or application placement. The present method differs by preserving a non-redundant broadcast objective: every healthy node receives the message once, and the recovery cost is measured by the number of external component-crossing repair edges.

The nearest same-objective line is local repair of a pruned broadcast tree. Rather than maintaining multiple complete trees in advance, the method studied here starts from one deterministic Gaussian broadcast tree, relocates the root to reduce fault damage, and then repairs the remaining components using the minimum possible number of external repair edges under the selected root. This gives a different tradeoff from independent spanning-tree protection: lower steady-state redundancy and very small repair-edge counts in common bounded-fault regimes, but no universal deterministic guarantee for arbitrary high-order clustered faults.

\section{Dense Gaussian Model and Broadcast Tree}
\label{sec:model}
Let $\Gk=(V,E)$ be the dense Gaussian network with
\begin{equation}
V=\Z_N,\qquad N=k^2+(k+1)^2.
\end{equation}
Two vertices $u,v\in V$ are adjacent if
\begin{equation}
v-u\equiv \pm k \quad\text{or}\quad v-u\equiv \pm(k+1) \pmod N.
\end{equation}
The coordinate map $\phi(x+yi)\equiv kx+(k+1)y\pmod N$ maps the Manhattan ball $\Ball$ bijectively onto $V$. Unit moves in the $x$ and $y$ directions correspond to the two generators $k$ and $k+1$.

For a root $r$, define $\Delta_r(v)=\modG(v-r)$ as the canonical coordinate difference from $r$ to $v$. A deterministic broadcast tree $T_r$ is formed by assigning each non-root node a parent that decreases $|x|+|y|$ by one in the coordinate difference. In the implementation used for the experiments, ties reduce the larger absolute coordinate first:
\begin{equation}
p(x,y)=
\begin{cases}
(x-\operatorname{sgn}(x),y), & |x|\ge |y|,~x\ne0,\\
(x,y-\operatorname{sgn}(y)), & \text{otherwise.}
\end{cases}
\end{equation}
The parent of $v$ is $r+\phi(p(x,y))$ modulo $N$.

\textbf{Lemma 1 (Fault-free tree).}
Every parent rule that decreases $|x|+|y|$ by one at each non-root node induces a spanning tree of $\Gk$ rooted at $r$ with depth at most $k$.

\emph{Proof.}
Every non-root node has one parent adjacent in $\Gk$, and the layer $|x|+|y|$ decreases by one. Repeated parent application reaches the root; hence directed cycles are impossible and every node is connected to the root. Since all canonical coordinates satisfy $|x|+|y|\le k$, the maximum depth is at most $k$.\hfill$\square$

\begin{figure}[H]
\centering
\includegraphics[width=0.6\linewidth]{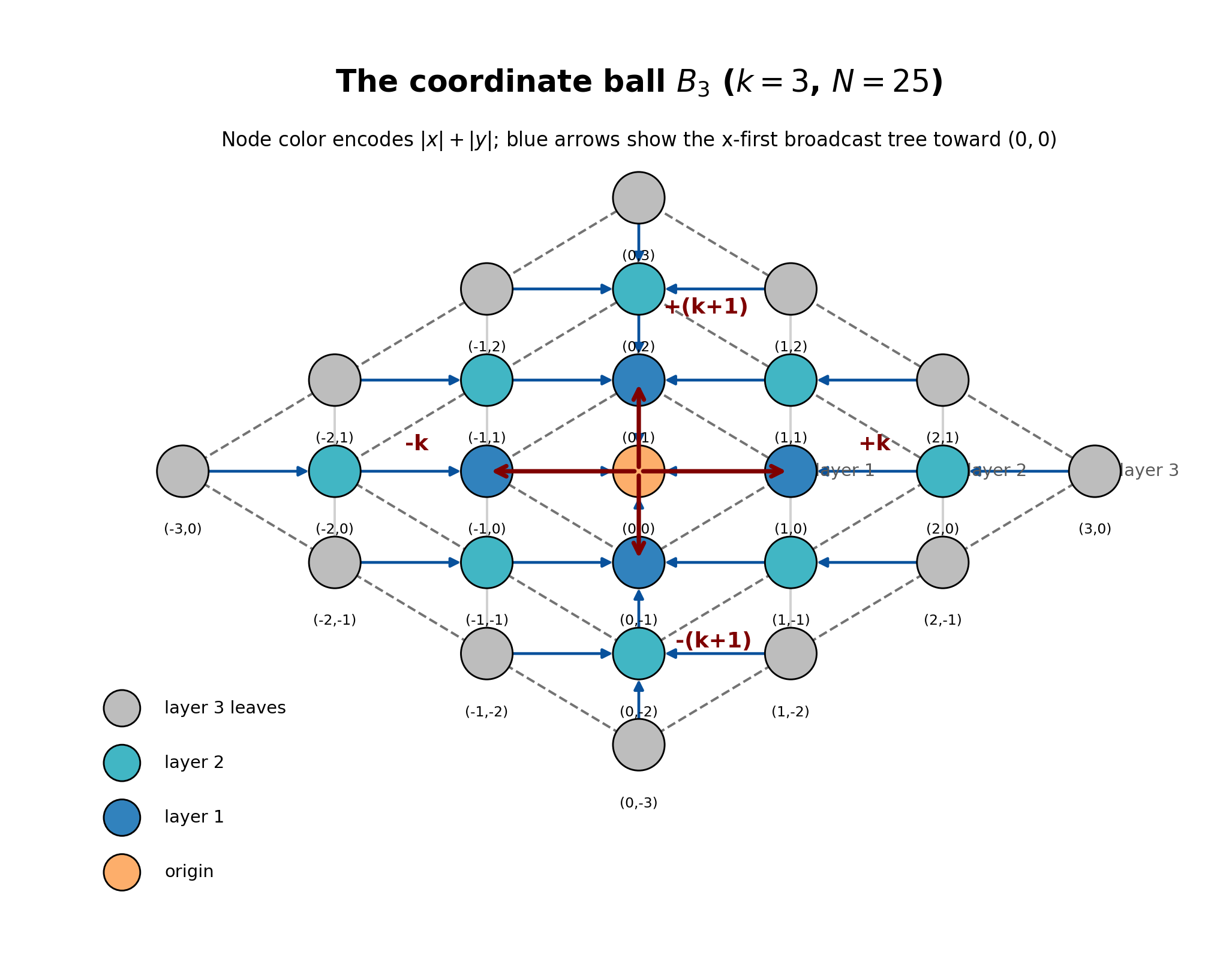}
\caption{The coordinate ball $B_3$ ($k=3$, $N=25$ nodes). Node color encodes layer $|x|+|y|$. Arrows show the x-first broadcast tree rooted at $(0,0)$. Layer-3 nodes are leaves and do not forward. The four generator directions correspond to adding or subtracting $k$ and $k+1$ in the integer label.}
\label{fig:bkdiagram}
\end{figure}

Figure~\ref{fig:bkdiagram} shows $B_3$ explicitly. Each node $(x,y)$ has exactly four Gaussian neighbors obtained by unit moves in the $x$ and $y$ directions, corresponding to adding or subtracting generator $k$ or $k+1$ modulo $N$. The tree assigns each non-root node the parent that reduces $|x|+|y|$ by one using the rule above. Boundary nodes at layer $k$ are leaves; this is the geometric basis for re-rooting.

\subsection{Recovery Objective}
Given a source $s$, node-fault set $\Fv$, and link-fault set $\Fe$, the recovery algorithm may select a root $r$ and a repaired tree $\widehat{T}_r$. The primary correctness objective is
\begin{equation}
V(\widehat{T}_r)=V\setminus\Fv,\qquad E(\widehat{T}_r)\cap\Fe=\emptyset,
\end{equation}
with exactly one parent for each healthy non-root vertex. The recovery objective has two forms. The experiments use the lexicographic score
\begin{equation}
\min (\rho_r, D_r, M_r),
\end{equation}
where $\rho_r$ is the number of external component-crossing repair edges, $D_r$ is the repaired depth, and $M_r$ is the number of candidate roots inspected. This ordering is appropriate for the control-state-sensitive setting studied here, because each repair edge creates an exceptional forwarding rule while small depth changes are already reported separately. It is a design choice, not a universal NoC objective: the paper prioritizes minimizing exceptional forwarding state and then reports depth separately. A latency-dominant implementation can instead use the weighted score
\begin{equation}
J_{\alpha,\beta,\gamma}(r)=\alpha \rho_r+\beta(D_r-k)+\gamma M_r,
\end{equation}
with $\beta>\alpha$ when hop latency dominates routing-state cost. Thus the lexicographic rule is one operating point in a family of state/latency tradeoffs. All tables therefore report both repair-edge count and depth so that another weighting can be applied without rerunning the correctness test.

\subsection{Fault Model}
Node faults are denoted by $\Fv\subseteq V$ and link faults by $\Fe\subseteq E$. The original source is assumed healthy. A valid recovered broadcast must reach every vertex in $V\setminus\Fv$ without using failed nodes or failed links. A link fault may be static, known before broadcast begins, or transient, discovered when a router attempts to forward across a failed neighbor. The evaluated transient model is deliberately limited to one runtime-discovered link fault followed by one recovery episode; transient node failures, cascading faults, and sequential multi-link discoveries are outside the evaluated scope and are listed as future work. Throughout the paper, \emph{re-rooting} denotes selecting a new effective broadcast source, while \emph{relocation} denotes the control movement or notification cost needed to activate that source.

Given a root $r$, pruning removes all faulty nodes, all tree edges incident to faulty nodes, and all failed tree links:
\begin{equation}
T_r^- = T_r - \Fv - \Fe.
\end{equation}
Let $\Comp_r=\{C_1,\ldots,C_c\}$ be the connected components of $T_r^-$. The healthy component graph $\Cgraph_r$ has one vertex for each $C_i$ and an edge $C_iC_j$ if there exists a healthy Gaussian edge in $\Gk-\Fv-\Fe$ with endpoints in $C_i$ and $C_j$.

\begin{figure}[H]
\centering
\includegraphics[width=0.85\linewidth]{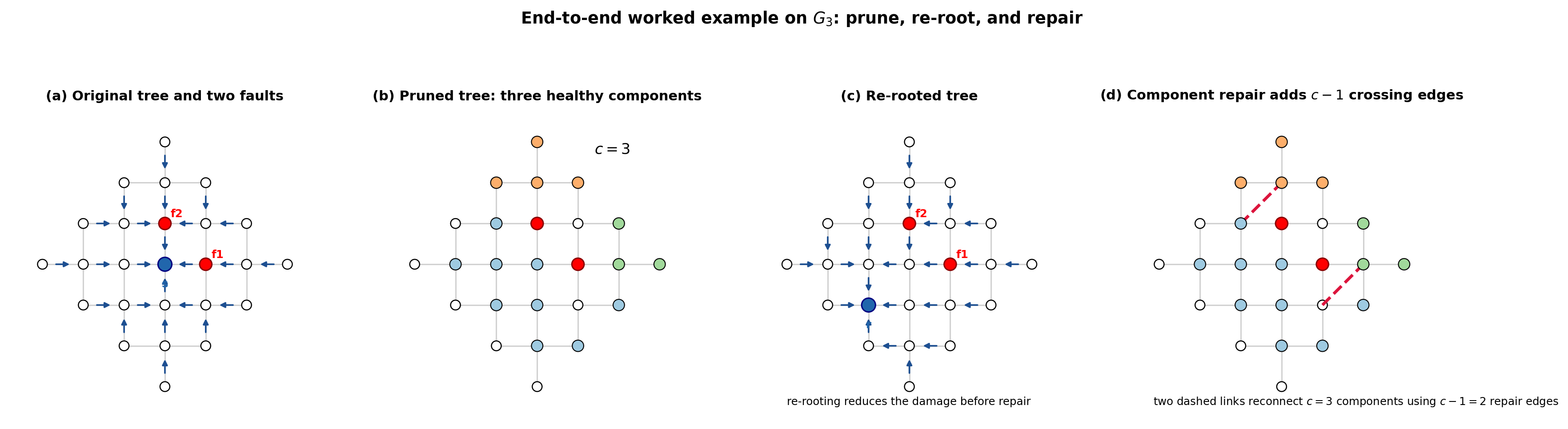}
\caption{Concrete end-to-end workflow example on $G_3$. Panel (a) shows the original tree rooted at the source with two failed nodes. Panel (b) shows the pruned tree split into $c=3$ healthy components. Panel (c) shows the re-rooted tree selected to reduce damage. Panel (d) shows the $c-1=2$ dashed repair edges that reconnect the components. This figure replaces the earlier abstract workflow sketch by a worked instance.}
\label{fig:workflow}
\end{figure}

\section{Re-Rooting Preliminaries}
\label{sec:reroot-prelim}
For node failures, the re-rooting goal is to choose a source $r$ satisfying
\begin{equation}
\dist(r,f)=k\qquad \forall f\in\Fv.
\end{equation}
Distance-$k$ vertices are leaves of a diameter-level broadcast tree, so they do not forward.

\textbf{Theorem 1 (Boundary intersection non-emptiness).}
For any two distinct nodes $a,b\in V(\Gk)$, the set
\begin{equation}
I(a,b)=\{r\in V:\dist(r,a)=k \text{ and } \dist(r,b)=k\}
\end{equation}
is nonempty. Moreover, $I(a,b)$ has at least two candidates for all non-degenerate fault pairs; in the finitely many degenerate tangent configurations it has exactly one candidate, which remains sufficient for Corollary~1.

\emph{Proof.}
By vertex transitivity it is enough to translate $a$ to the origin and write $C=b-a$ in canonical coordinates. A root $r$ is valid exactly when its offset $P=r-a$ satisfies
\begin{equation}
P\in \partial B_k,\qquad P-C\in \partial B_k
\end{equation}
in one of the quotient copies of the coordinate ball. Equivalently, $C$ must be expressible as a difference of two boundary points of the quotient ball. The dense Gaussian quotient has lattice generators $(k,k+1)$ and $(-(k+1),k)$. Hence it is sufficient to inspect the central copy and the eight adjacent copies, i.e., shifts $m(k,k+1)+n(-(k+1),k)$ for $m,n\in\{-1,0,1\}$. On each shifted copy, the equations
\begin{equation}
|p|+|q|=k,
\qquad
|p-c'|+|q-d'|=k
\end{equation}
are the intersection equations for two Manhattan diamonds. Removing absolute values gives the sixteen signed systems in Section~\ref{sec:candidate}. The side-pair cases are exhaustive because every boundary point lies on one of four sides and every shifted copy contributes four translated sides. If the two sides are nonparallel, the signed linear system has one candidate; if they are parallel, the intersection is either empty or an interval endpoint is obtained by the same side inequalities. The dense quotient identity $\partial B_k-\partial B_k=B_k$ follows from this finite side-pair enumeration: every canonical difference $C\in B_k$ appears in at least one of the nine shifted side pairs, guaranteeing that $I(a,b)$ is nonempty.

For the ``at least two candidates'' claim, three intersection cases arise for a given shifted side pair. Case 1 is nonparallel sides with transverse crossing. The two signed linear systems have determinant nonzero and the solution $(p,q)$ lies in the interior of both boundary sides. The opposite sign choice $(-s_1,-s_2,s_3,s_4)$ or $(s_1,s_2,-s_3,-s_4)$ yields a second solution symmetric about the side midpoint. These two solutions are distinct because $C\ne0$ prevents the symmetric point from coinciding with the original.

Case 2 is parallel sides with interval intersection. The two boundary sides are parallel and their overlap is a line segment. The two endpoints of this segment are distinct candidates whenever the overlap has positive length. Zero-length overlap means that the sides are tangent at one point; this is the degenerate case.

Case 3 is tangent intersection. Tangency occurs when $C$ lies exactly on a boundary side of a shifted copy with the same slope as that side, so the two diamonds meet at exactly one point. In this case both sign orientations yield the same candidate. There is therefore one valid candidate, not two.

Cases 1 and 2 with positive-length overlap yield at least two candidates. Case 3 yields exactly one. Since Case 3 occurs only on a finite set of boundary configurations defined by the nine shifted boundary sides, it is a degenerate condition. Thus $I(a,b)$ is nonempty for all nonzero canonical differences, and it has at least two candidates except in the finitely many degenerate tangent configurations.\hfill$\square$

\textbf{Corollary 1 (One/two node re-rooting primitive).}
For every one- or two-node fault set $\Fv$ in $\Gk$, there exists a root $r$ such that $\dist(r,f)=k$ for all $f\in\Fv$.

\emph{Proof.}
For one fault, choose any boundary vertex of the failed node. For two faults $a,b$, apply Theorem~1 and select any $r\in I(a,b)$.\hfill$\square$

\textbf{Corollary 2 (Zero-repair node cases).}
If $|\Fv|\le2$, $\Fe=\emptyset$, and $r$ satisfies $\dist(r,f)=k$ for every failed node, then $T_r^-$ has one healthy component and requires zero repair edges.

\emph{Proof.}
The failed nodes are leaves in $T_r$. Removing leaves cannot disconnect healthy internal forwarding nodes from $r$. Thus $c=1$.\hfill$\square$

\begin{figure}[H]
\centering
\includegraphics[width=0.6\linewidth]{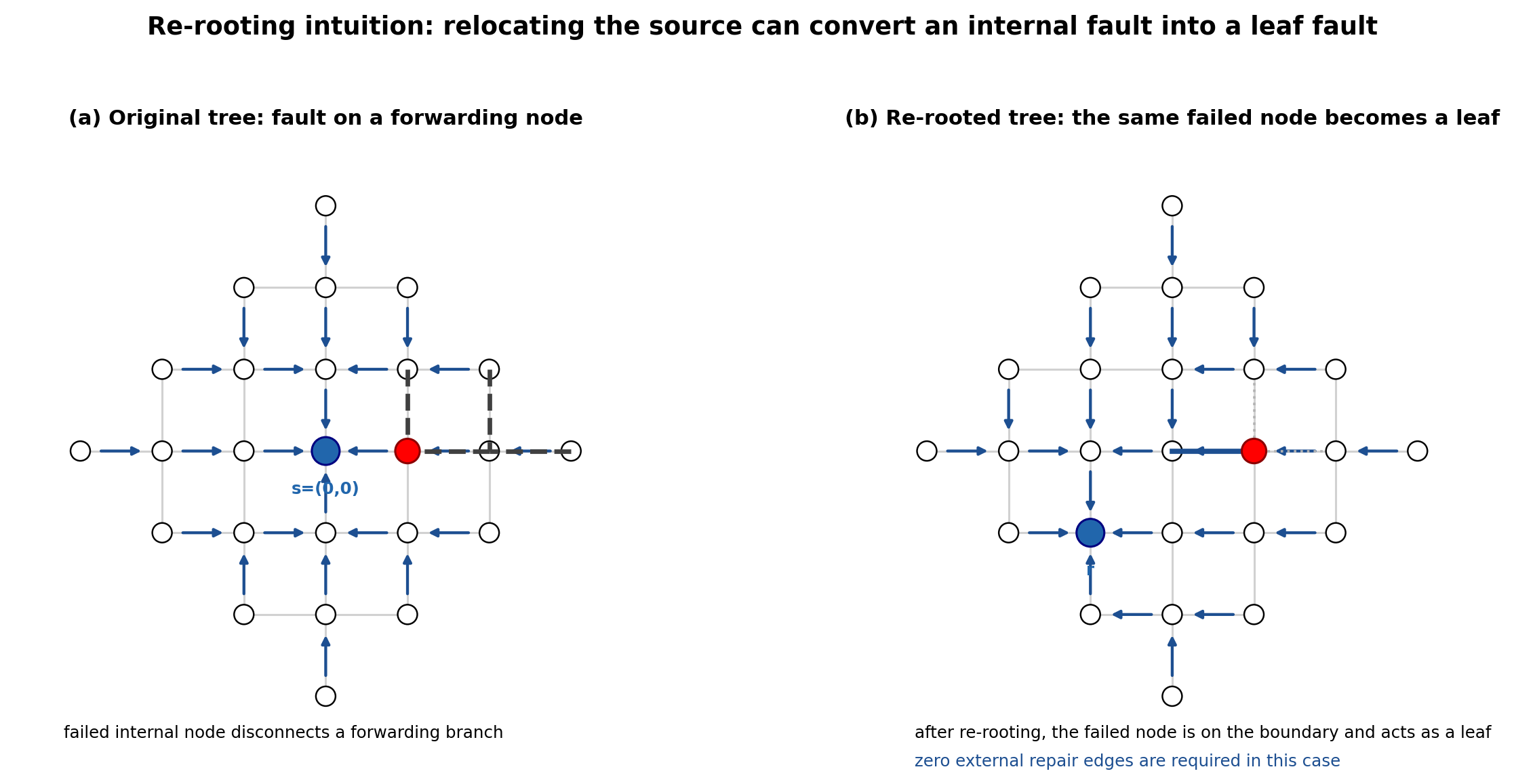}
\caption{Two-panel re-rooting walkthrough. Left: in the original tree, a failed internal node disconnects a forwarding branch. Right: after re-rooting, the same failed node is on the boundary and behaves as a leaf, so the tree remains connected and zero repair edges are needed.}
\label{fig:reroot_walkthrough}
\end{figure}

\textbf{Example 1 (zero-repair re-rooting).}
Consider $k=3$ with failed nodes at $(2,1)$ and $(-1,2)$. With the original source at $(0,0)$, both faults may appear on forwarding branches. Re-rooting to the illustrated source moves those faults to the boundary of the new broadcast tree, where they act as leaves. The pruned tree remains connected, so zero external repair edges are required. Figure~\ref{fig:reroot_walkthrough} gives the core intuition in a small worked example.

\section{Constant-Size Candidate Generation}
\label{sec:candidate}
This section gives the compact candidate-generation primitive used as the hardware-oriented selector. It is included so the present paper is not dependent on a separate constant-time selector manuscript.

Let $C=(c,d)=\modG(b-a)$ be the reduced difference of two failed nodes. A valid offset $P=(p,q)$ satisfies
\begin{equation}
|p|+|q|=k,
\qquad
|p-c'|+|q-d'|=k,
\label{eq:shiftedabs}
\end{equation}
for some quotient-lattice shift
\begin{equation}
(c',d')=(c,d)+m(k,k+1)+n(-(k+1),k),
\end{equation}
where $m,n\in\{-1,0,1\}$. The nine shifts cover the central copy and the eight adjacent quotient copies that can intersect the canonical boundary.

For each shift, remove the absolute values in~\eqref{eq:shiftedabs} by choosing signs
\begin{equation}
(s_1,s_2,s_3,s_4)\in\{\pm1\}^4.
\end{equation}
The signed system is
\begin{equation}
s_1p+s_2q=k,
\qquad
s_3p+s_4q=k+s_3c'+s_4d'.
\label{eq:signsystem}
\end{equation}
When the determinant is nonzero, Cramer's rule gives one candidate. When the equations are parallel, the valid candidates form an interval on a boundary side, and an endpoint can be selected. Hence at most $9\times16=144$ signed systems are checked.

\textbf{Lemma 2 (Constant-size two-node generator).}
Under the standard word-RAM model for coordinate arithmetic, the shifted sign-case procedure returns a valid two-node re-rooting candidate in $O(1)$ arithmetic cases.

\emph{Proof.}
The number of quotient shifts is nine and the number of sign tuples per shift is sixteen. Each nonparallel case requires constant arithmetic; each parallel case requires constant interval endpoint checks. Existence follows from the boundary-intersection theorem in Section~\ref{sec:reroot-prelim}.\hfill$\square$

\subsection{Selector Used in the Experiments}
The validation code intentionally separates the mathematical selector from implementation choices. For one and two node faults, the experiments used exact boundary-intersection enumeration of the $4k$ boundary candidates rather than the compressed $144$-case selector. This makes the reported recovery independent of any possible implementation bias in the constant-size selector; the constant-size selector is included as the deployable primitive, while enumeration is used as a conservative validator. This gives $O(k)$ candidate generation for the node-guaranteed cases. For three or more node faults, candidates are ranked by the number of failed nodes placed at distance $k$. For link-only cases, candidate roots are tried in a deterministic pseudo-random order with a cap of 20,000 roots in the full run; this cap is used only for the avoid-only selector. The hybrid repair still prunes and repairs the selected tree even when no perfectly link-avoiding root is found. Link-safety testing for a candidate $r$ costs $O(|\Fe|)$ because each failed link $e=(u,v)$ is checked by testing whether the parent of $u$ is $v$ or the parent of $v$ is $u$ in $T_r$.

Thus the experimental source-selection cost is explicitly
\begin{equation}
O(M(|\Fv|+|\Fe|)),
\end{equation}
where $M$ is the number of candidate roots checked. For one/two node faults, $M\le4k$; for full-run link-only avoidance, $M\le20{,}000$; for the constant-size selector, $M\le144$. The component-repair phase scans the $N$ nodes and their four Gaussian neighbors, giving $O(N)$ time for a fixed selected root. No experimental setting uses an $O(N^2)$ repair step.

\begin{table}[H]
\centering
\caption{Candidate-generation strategies and their role.}
\label{tab:candidate_complexity}
\footnotesize
\begin{tabular}{p{0.17\linewidth}p{0.34\linewidth}p{0.14\linewidth}p{0.20\linewidth}}
\toprule
Regime & Candidate rule & Cases & Purpose \\
\midrule
1 node & boundary around fault & $4k$ & exact validation \\
2 nodes & boundary intersection & $\le4k$ & exact validation \\
2 nodes & shifted sign systems & $\le144$ & hardware primitive \\
$\ge3$ nodes & max leaf-score ranking & bounded & heuristic \\
link-only & ordered root trials & $\le 20000$ & avoid-only comp. \\
repair & component scan & $O(N)$ & final recovery \\
\bottomrule
\end{tabular}
\end{table}

\subsection{Constant-Size Selector Cross-Validation}
The constant-size selector and the $O(k)$ boundary enumeration are allowed to return different roots; the correctness requirement is that the returned root satisfy the same distance-$k$ predicate. To close the implementation gap between the hardware-oriented primitive and the conservative validator, the selector was cross-checked against exact boundary enumeration on the same one- and two-node instances used in the full run. Table~\ref{tab:selector_crosscheck} reports zero misses and zero invalid roots. Thus the experimental use of enumeration does not hide a correctness problem in the constant-size primitive; enumeration was used to make the validation independent of sign-case implementation details.

\begin{table}[H]
\centering
\caption{Cross-validation of the $9\times16$ constant-size selector against exact boundary enumeration.}
\label{tab:selector_crosscheck}
\begin{tabular}{lrrrr}
\toprule
Fault regime & Trials & Valid roots & Misses & Invalid outputs \\
\midrule
One node & 20,000 & 20,000 & 0 & 0 \\
Two nodes & 20,000 & 20,000 & 0 & 0 \\
All one/two-node cases & 40,000 & 40,000 & 0 & 0 \\
\bottomrule
\end{tabular}
\end{table}

\subsection{Link-Safety Candidate Structure}
For node faults, the boundary-intersection structure reduces candidate search to a constant number of shifted side cases. An analogous reduction for link faults would require characterizing all roots $r$ for which $\lambda(r,\Fe)=0$, that is, roots whose broadcast tree avoids every failed link. A failed link $e=(u,v)$ is avoided by $T_r$ if and only if neither $u$ is the parent of $v$ nor $v$ is the parent of $u$ under the coordinate rule rooted at $r$. The parent relation depends on the two relative coordinates $\Delta_r(u)$ and $\Delta_r(v)$ simultaneously, so the resulting constraint is a region in root-coordinate space rather than a single boundary-distance equation. It does not decompose as cleanly as the distance-$k$ node-fault condition.

\textbf{Proposition 3 (Link-avoidance exclusion condition).}
Let $e=\{u,v\}$ be a failed Gaussian link and let $r$ be a candidate root. Write
\begin{equation}
\Delta_r(w)=\modG(w-r)
\end{equation}
for the canonical coordinate of node $w$ relative to root $r$, and let $p(\cdot)$ be the coordinate parent rule in (7). Then $T_r$ uses the failed link $e$ if and only if
\begin{equation}
\Delta_r(u)=p(\Delta_r(v))
\quad\text{or}\quad
\Delta_r(v)=p(\Delta_r(u)).
\label{eq:link_exclusion}
\end{equation}
Equivalently, if $\Delta_r(v)=(x,y)$, the first equality in (\ref{eq:link_exclusion}) holds exactly when one of the following parent-reduction cases holds:
\begin{align}
\Delta_r(u)&=(x-\operatorname{sgn}(x),y), && |x|\ge |y|,~x\ne0,\label{eq:link_x}\\
\Delta_r(u)&=(x,y-\operatorname{sgn}(y)), && \text{otherwise and }y\ne0.\label{eq:link_y}
\end{align}
The two symmetric cases are obtained by exchanging $u$ and $v$.

\emph{Proof.}
By definition, the broadcast tree rooted at $r$ assigns each non-root vertex $w$ the parent whose coordinate is $p(\Delta_r(w))$. Since $\phi$ is a bijection from the coordinate ball to the vertex set, the parent of $v$ is $u$ if and only if $\Delta_r(u)=p(\Delta_r(v))$. Similarly, the parent of $u$ is $v$ if and only if $\Delta_r(v)=p(\Delta_r(u))$. Expanding the two branches of the parent rule gives (\ref{eq:link_x}) and (\ref{eq:link_y}). Therefore the undirected failed link is used by $T_r$ exactly in the two cases stated in (\ref{eq:link_exclusion}).\hfill$\square$

\textbf{Corollary 4 (Link-avoidance root characterization).}
For a fixed failed link $e=\{u,v\}$, define
\begin{equation}
A(e)=\{r\in V: \Delta_r(u)=p(\Delta_r(v))\text{ or }\Delta_r(v)=p(\Delta_r(u))\}.
\end{equation}
Then a candidate root $r$ avoids $e$ if and only if $r\notin A(e)$. For a failed-link set $F_E$, a root is link-safe if and only if
\begin{equation}
r\notin \bigcup_{e\in F_E} A(e).
\end{equation}

Proposition~3 gives a necessary and sufficient $O(1)$ coordinate test for whether a candidate root uses a specific failed link; it does not require constructing the full tree. For multiple failed links, candidate testing costs $O(|F_E|)$ coordinate comparisons. The set $A(e)$ is a structured exclusion set in root-coordinate space. This suggests that a closed-form link selector may be possible by intersecting the node-boundary candidate system with these link-exclusion sets, but a complete constant-size construction for arbitrary link-fault sets remains open.

A complete constant-size link-avoidance selector remains open. The present experiments therefore use a deterministic cap-bounded root order for the avoid-only comparator. This is sufficient for the hybrid method because perfect link avoidance is not required for correctness: residual failed tree links are converted into component cuts and repaired by Theorem~2 whenever the component graph is connected. Table~\ref{tab:link_candidate_overhead} shows that the average candidates checked remain far below the cap for one to three failed links; cap hits occur only in the five-link stress regime already labeled empirical.

\section{Link-Failure Exclusion and Component Repair}
\label{sec:link-repair}
For a candidate root $r$, define the failed-tree-link count
\begin{equation}
\lambda(r,\Fe)=|E(T_r)\cap\Fe|.
\end{equation}
If $\lambda(r,\Fe)=0$, the selected tree avoids all failed links. If $\lambda(r,\Fe)>0$, the failed links split the tree into components that may still be reconnectable by healthy Gaussian edges.

\textbf{Lemma 3 (Repair lower bound).}
If $T_r^-$ has $c$ healthy components, then any non-redundant repaired broadcast tree preserving these internal components must use at least $c-1$ external component-crossing repair edges.

\emph{Proof.}
Contract each healthy tree component to a supernode. Any connected repaired broadcast must connect the $c$ supernodes, and any connected graph on $c$ vertices has at least $c-1$ edges.\hfill$\square$

\textbf{Theorem 2 (Edge-minimum component repair).}
If the healthy component graph $\Cgraph_r$ is connected, then $T_r^-$ can be repaired into a non-redundant broadcast tree over all healthy nodes using exactly $c-1$ external component-crossing repair edges. This is minimum.

\emph{Proof.}
Since $\Cgraph_r$ is connected, choose a spanning tree of $\Cgraph_r$ rooted at the component containing $r$. For every non-root component, add one corresponding healthy Gaussian crossing edge. The internal components are trees because they are subgraphs of $T_r$. Contracting every component maps the repaired structure to the selected component-level spanning tree; hence no cycle is introduced. The final graph is connected, acyclic, spans all healthy vertices, and can be rooted at $r$. It uses $c-1$ repair edges, which is optimal by Lemma~3.\hfill$\square$

\textbf{Lemma 4 (Component-count bound and depth accounting).}
Let $q_E^T=|E(T_r)\cap\Fe|$ be the number of failed links used by the selected tree, and let $\ell_T$ be the number of failed nodes that are leaves of $T_r$. Then pruning creates at most
\begin{equation}
 c \le 1+q_E^T+3(|\Fv|-\ell_T)
\label{eq:component_bound}
\end{equation}
healthy components. Consequently, if the component graph is connected, the edge-minimum repaired tree uses
\begin{equation}
R=c-1\le q_E^T+3(|\Fv|-\ell_T)
\label{eq:repair_bound}
\end{equation}
external component-crossing repair edges. The hop depth of the repaired tree is a separate measured quantity: it depends on the endpoints selected for the component-crossing edges and on the rooted orientation of the repaired tree, and is therefore not bounded in general by $k+R$.

\emph{Proof.}
Deleting a tree link increases the number of components by at most one, giving the $q_E^T$ term. A failed leaf contributes no new healthy component because it has no child subtree. A non-leaf failed node in the coordinate-reduction tree has at most one parent and at most two child subtrees under the binary coordinate-priority rule used here; deleting it can therefore increase the number of healthy components by at most three. Summing these contributions gives~\eqref{eq:component_bound}. If the component graph is connected, Theorem~2 selects exactly $c-1$ external component-crossing edges, which gives~\eqref{eq:repair_bound}. The depth statement is an accounting clarification rather than an additional bound: after components are reconnected, a root-to-destination path may enter a component through a repair endpoint that is not the original parent-side gate of that component, and the path may then traverse a nontrivial internal subtree segment. Thus a single repair edge can add more than one hop to the longest rooted path, even though it contributes only one exceptional forwarding rule. This is why the experiments report repaired depth separately from $R$.\hfill$\square$

\textbf{Proposition 5 (Hybrid repaired-depth accounting).}
If the selected hybrid root yields $R=0$, the repaired structure is the pruned re-rooted Gaussian broadcast tree, so its depth is at most $k$ by Lemma~1 and is exactly $k$ whenever a healthy boundary vertex remains. If $R>0$ and the repair reconnects components $C_1,\ldots,C_c$, then a coarse accounting bound is $D_r\le k+\sum_i \operatorname{diam}(C_i)$, with $\operatorname{diam}(C_i)\le 2k$ because each component is a subtree of a depth-$k$ Gaussian tree. This bound is loose but resolves the interpretation: $R$ counts exceptional component-crossing forwarding rules, whereas depth also depends on the internal distance from a repair-entry gate to the farthest descendant inside each reconnected component. The near-$k$ hybrid depths in the experiments therefore reflect the observed small residual component diameters after re-rooting, not a universal $k+R$ bound.\hfill$\square$

\textbf{Proposition 1 (Component-count bound tightness).}
The bound in~\eqref{eq:component_bound} is worst-case tight when every non-leaf failed node has one parent edge and two child subtrees in $T_r$, and when all failed tree links lie on distinct tree paths.

\emph{Proof.}
In such a placement, each failed tree link separates one previously connected child subtree, increasing $c$ by exactly one. Each non-leaf failed node with one parent and two child subtrees separates the parent-side component and two child-side components, increasing the count by exactly three relative to the intact tree. If the affected paths are distinct, none of these increases merge or overlap. The upper bound is therefore attained.\hfill$\square$

The bound is deliberately a worst-case tool. Boundary faults have fewer children, and multiple faults often lie on the same subtree path. Hence practical component counts are much lower than the bound for node faults; Table~\ref{tab:component_bound_eval} quantifies this gap.

\begin{figure}[H]
\centering
\includegraphics[width=0.9\linewidth]{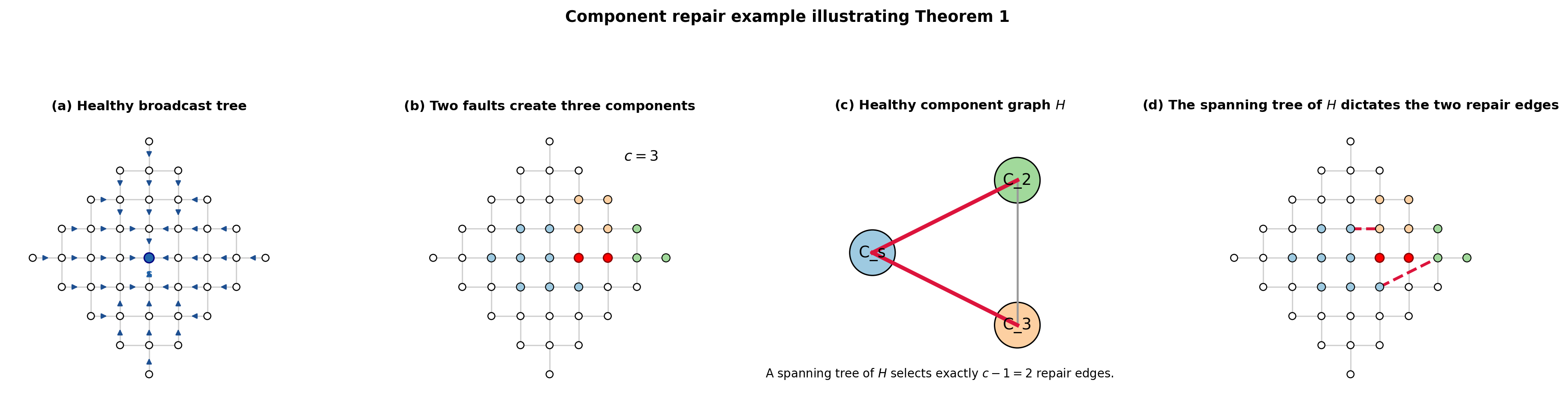}
\caption{Component-repair example illustrating Theorem~2. Panel (a) shows a healthy tree. Panel (b) shows the tree after pruning two faults, creating three healthy components. Panel (c) draws the healthy component graph $H$ explicitly and highlights a spanning tree of $H$. Panel (d) maps that spanning tree back to exactly $c-1=2$ component-crossing repair edges in the original network.}
\label{fig:components}
\end{figure}

\textbf{Example 3 (multi-component repair).}
Figure~\ref{fig:components} provides a worked example with $c=3$ components. The healthy component graph $H$ has three vertices and a highlighted spanning tree. Mapping those two highlighted component-graph edges back to the Gaussian network gives exactly the two dashed repair edges in panel (d), making the statement of Theorem~2 visually immediate.

\textbf{Lemma 5 (Healthy graph implies component graph).}
If the healthy graph $\Gk-\Fv-\Fe$ is connected, then $\Cgraph_r$ is connected for every pruned tree $T_r^-$ over the healthy vertices.

\emph{Proof.}
For any two tree components, a path in the connected healthy graph joins them. Whenever this path moves from one tree component to another, it uses a healthy crossing edge; therefore the path induces a path in $\Cgraph_r$.\hfill$\square$

\textbf{Lemma 6 (No Gaussian edge is a bridge).}
Every edge of $\Gk$ lies on a 4-cycle. Therefore deleting one link does not disconnect $\Gk$.

\emph{Proof.}
For an edge $(u,u+k)$, the vertices
\begin{equation}
u,
\quad u+k,
\quad u+k+(k+1),
\quad u+(k+1)
\end{equation}
form a 4-cycle using generator differences $k,k+1,-k,-(k+1)$. The generator $k+1$ case is symmetric.\hfill$\square$

\textbf{Theorem 3 (Deterministic single-link repair).}
For any one failed link $e\in E(\Gk)$ and any healthy selected root $r$, the hybrid repair phase succeeds. If $e\notin E(T_r)$, zero repair edges are needed. If $e\in E(T_r)$, exactly one repair edge is necessary and sufficient.

\emph{Proof.}
If $e$ is not in $T_r$, pruning does not change the tree. If $e$ is in $T_r$, deleting it splits the tree into exactly two components. By Lemma~6, $\Gk-e$ is connected; by Lemma~5, the two-component graph is connected. Theorem~2 repairs it with $c-1=1$ edge, and Lemma~3 proves that one edge is necessary.\hfill$\square$

\textbf{Proposition 2 (Minimum isolating cut).}
For any healthy node $v$ in $\Gk$, the minimum node-fault set that isolates $v$ from all other healthy nodes has size four, consisting of the four Gaussian neighbors of $v$. The minimum link-fault set that isolates $v$ has size four, consisting of the four incident edges of $v$.

\emph{Proof.}
Every vertex of $\Gk$ has exactly four incident Gaussian channels, corresponding to the generators $\pm k$ and $\pm(k+1)$. To isolate a fixed healthy node $v$ as a singleton component, every incident channel from $v$ to a healthy neighbor must be removed. With node faults this requires faulting all four Gaussian neighbors of $v$; any set of at most three faulty neighbors leaves at least one healthy neighbor adjacent to $v$. Thus four node faults are necessary and the four-neighbor set is sufficient. The same local argument applies to links: all four incident edges of $v$ must be removed, and the four incident edges are sufficient.\hfill$\square$

\textbf{Example 2 (single-link repair).}
Take $k=3$ and let the tree edge between $(1,0)$ and $(2,0)$ fail. The local 4-cycle
\[(1,0)\rightarrow(2,0)\rightarrow(2,1)\rightarrow(1,1)\rightarrow(1,0)\]
contains the failed edge and provides a healthy bypass across the top side of the cycle. Thus a single crossing edge reconnects the two pruned tree components.

\begin{figure}[H]
\centering
\includegraphics[width=0.6\linewidth]{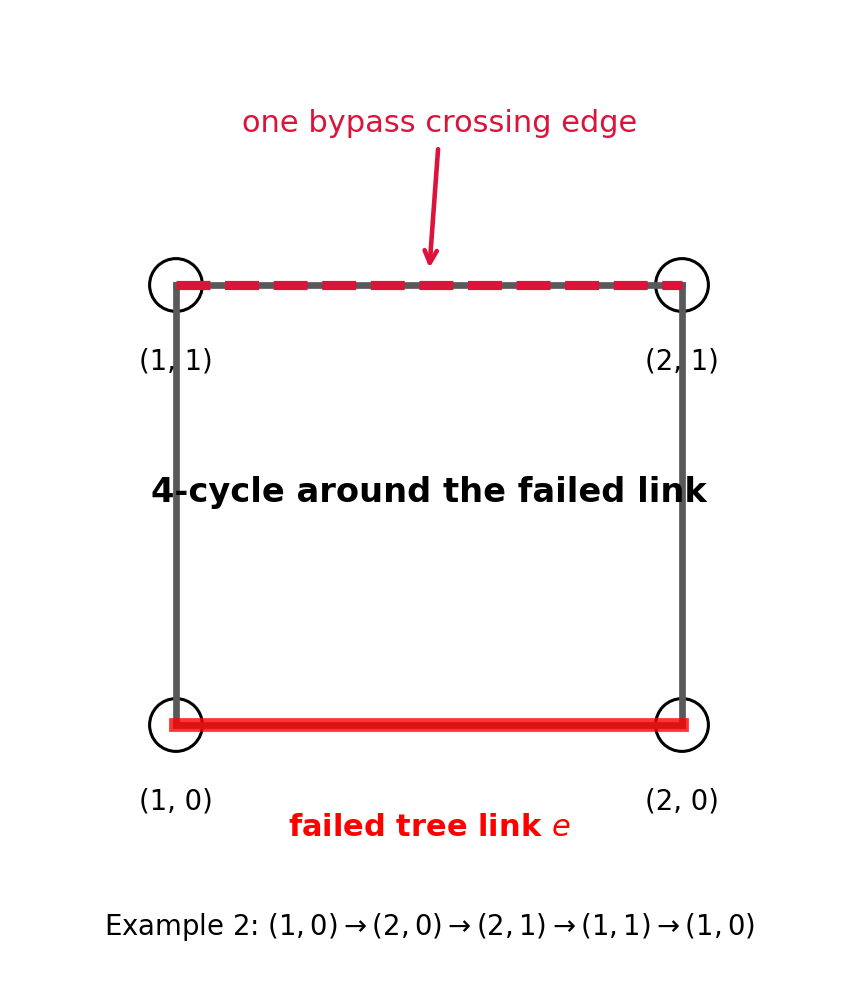}
\caption{Single-link fault bypass for Lemma~6 and Theorem~3. The failed tree link lies on a 4-cycle, so the two resulting components are reconnected by exactly one crossing edge.}
\label{fig:link4cycle}
\end{figure}

\section{Runtime Micro-Re-Rooting Protocol}
\label{sec:runtime}
The runtime protocol handles a failed link discovered during broadcast propagation. Suppose node $u$ attempts to forward a packet to child $v$ and detects that edge $(u,v)$ is unavailable. The detector reports the failed link and its current layer. The recovery action has three logical phases.

\textbf{Phase 1: local observation.}
The detector records the failed link and the partially reached set. In a strictly local implementation, the affected subtree can be treated as the recovery region. In the present experiments, we evaluate the conservative global recovery variant: the detector triggers a re-rooting and repair over the whole broadcast instance. This gives an upper bound on recovery scope and isolates the source-selection and repair behavior.

\textbf{Phase 2: candidate selection and link test.}
The candidate generator returns one or more roots. Each candidate is tested by $\lambda(r,\Fe)$, the number of failed links used by its tree. A root with $\lambda=0$ is preferred. If none is found, the root minimizing failed tree links and maximizing node-leaf score is used.

\textbf{Phase 3: component repair.}
The selected tree is pruned, the healthy components are contracted, and a component spanning tree is selected if $\Cgraph_r$ is connected. The detector-to-root relocation is routed through the healthy graph. If relocation is impossible, recovery fails; otherwise, the repaired tree is used to complete delivery.

The transient protocol therefore supports both a global form, used in the experiments, and a local/regional form in which $\Cgraph_r$ is built only for the affected subtree. The local form is the natural hardware optimization because only descendants of the detected failed forwarding edge need to be reconsidered. The global form is the safer validation model because it does not rely on optimistic assumptions about regional containment. Thus the reported transient latency is best interpreted as a conservative upper-level recovery cost rather than a claim that every implementation must pause the whole NoC.

\textbf{Example 4 (transient discovery).}
Suppose the propagation front has reached layer $\ell=7$ when a router discovers that its next forwarding link is unavailable. The detector reports the failed link, the source is relocated by $d(s,r)$ hops, and the repaired tree then finishes in depth $D_{\mathrm{repaired}}$. The recovery time decomposition
\begin{equation}
T_{\mathrm{recover}}=\ell_{\mathrm{detect}}+d(s,r)+D_{\mathrm{repaired}}
\end{equation}
is reported later through the transient discovery-layer table, where three placement classes exhibit different discovery times but the same logical three-phase structure.

\subsection{Regional Recovery Cost Bound}
The global transient experiment is a conservative upper-level model. If a failed tree edge is discovered at layer $\ell$, then the affected downstream region has remaining height at most $h=k-\ell$. A Gaussian ball of radius $h$ contains $1+2h(h+1)$ vertices, so a regional repair that scans only the affected subtree has the bound
\begin{equation}
N_{\mathrm{local}}(\ell)\le 1+2(k-\ell)(k-\ell+1).
\end{equation}
Relative to $N=1+2k(k+1)$, the scan fraction is at most
\begin{equation}
\frac{N_{\mathrm{local}}(\ell)}{N}\le
\frac{1+2(k-\ell)(k-\ell+1)}{1+2k(k+1)}.
\end{equation}
For $k=200$, the measured discovery layers imply the regional upper bounds in Table~\ref{tab:regional_bound}. Boundary-discovered faults require almost no regional scan, random tree-edge failures require about 11.4\% of the graph on average, and only near-root failures approach the global cost.

\begin{table}[H]
\centering
\caption{Estimated regional scan bound for transient recovery at $k=200$.}
\label{tab:regional_bound}
\begin{tabular}{lrrr}
\toprule
Placement & Avg. $\ell$ & Avg. $h$ & Local scan bound \\
\midrule
Critical near root & 34.56 & 165.44 & 68.50\% \\
Random tree edge & 132.79 & 67.21 & 11.40\% \\
Boundary branch & 199.52 & 0.48 & 0.003\% \\
\bottomrule
\end{tabular}
\end{table}

\section{Higher-Order Best-Effort Heuristic}
\label{sec:heuristic}
For three or more node faults, a common distance-$k$ source is not guaranteed. Therefore high-order recovery is explicitly treated as best effort. For a candidate root $r$, define the leaf score
\begin{equation}
L(r)=|\{f\in\Fv:\dist(r,f)=k\}|.
\end{equation}
The heuristic ranks roots lexicographically by high $L(r)$ and low $\lambda(r,\Fe)$. After a root is selected, the repair theorem is still exact: if $\Cgraph_r$ is connected, the repair uses the minimum $c-1$ edges. What becomes heuristic is the existence and selection of a root that keeps $\Cgraph_r$ connected in arbitrary high-order clustered fault patterns.

\textbf{Theorem 4 (Re-rooting reliability threshold and triple-intersection criterion).}
Let $F_V$ be a node-fault set in $\Gk$.
For $|F_V|\le2$, a root $r$ satisfying $\dist(r,f)=k$ for every $f\in F_V$ always exists, and if $F_E=\emptyset$ the repaired broadcast requires zero component-crossing repair edges.
For $|F_V|=3$, a zero-repair re-rooting root exists if and only if
\begin{equation}
\bigcap_{f\in F_V}\{r\in V:\dist(r,f)=k\}\ne\emptyset.
\label{eq:triple_intersection}
\end{equation}
Equivalently, for $F_V=\{f_1,f_2,f_3\}$, this condition is
\begin{equation}
I(f_1,f_2)\cap I(f_1,f_3)\ne\emptyset.
\end{equation}
There is no universal three-node guarantee: for example, in $G_3$ the three faults $(-3,0)$, $(-1,0)$, and $(1,0)$ have no common distance-three root.

\emph{Proof.}
The $|F_V|\le2$ case is Corollaries~1 and~2. For three faults, a root gives zero-repair re-rooting precisely when every failed node is at layer $k$ relative to that root. This is exactly the intersection condition (\ref{eq:triple_intersection}). For $F_V=\{f_1,f_2,f_3\}$, the first two constraints define $I(f_1,f_2)$ and adding the third constraint is equivalent to intersecting with $I(f_1,f_3)$. The $G_3$ example can be checked directly on the 25-node coordinate ball: no coordinate $(x,y)$ with $|x|+|y|\le3$ has Manhattan distance three from all of $(-3,0)$, $(-1,0)$, and $(1,0)$ under the dense Gaussian quotient. Hence the one/two-fault guarantee is the deterministic threshold for universal zero-repair re-rooting, while three-fault success depends on the triple-intersection geometry.\hfill$\square$

Theorem~4 gives an exact test for three-node zero-repair re-rooting but intentionally does not claim that all three-fault sets are recoverable by re-rooting alone. The hybrid method can still recover many three-fault cases after component repair, but this recovery is conditional on the selected component graph remaining connected.

\textbf{Proposition 4 (Local-obstruction probability bound).}
Let $q$ node faults be sampled uniformly without replacement from $V(\Gk)$, and let $\mathcal{P}_k$ be any family of local obstruction patterns such that $|\mathcal{P}_k|\le C N$ and every pattern consists of $s$ specified faulty vertices, where $C$ and $s$ are independent of $k$. Then
\begin{align}
&\Pr[\exists P\in\mathcal{P}_k:\; P\subseteq F_V] \notag\\
&\qquad\le C N \left(\frac{q}{N}\right)^s
= O\!\left(\frac{q^s}{k^{2s-2}}\right).
\end{align}
In particular, the four-neighbor isolation obstruction of Proposition~2 has $s=4$ and probability $O(q^4/k^6)$ for fixed $q$.

\emph{Proof.}
For any fixed pattern $P$ of size $s$, the probability that all vertices of $P$ are sampled is
\begin{equation}
\frac{\binom{N-s}{q-s}}{\binom{N}{q}}\le \left(\frac{q}{N}\right)^s.
\end{equation}
A union bound over at most $C N$ translated local patterns gives the first inequality. Since $N=2k^2+2k+1=\Theta(k^2)$, the asymptotic form follows. For four-neighbor isolation, $s=4$ and there is one translated pattern per vertex up to a constant factor.\hfill$\square$

Proposition~4 is not a complete guarantee that $\Cgraph_r$ is connected; component-graph disconnection can occur through more global arrangements than a single isolated vertex. It does, however, formalize the density effect observed in Table~\ref{tab:rootcause}: fixed-size local obstructions become rapidly less likely as $k$ grows, while close-cluster placements deliberately concentrate faults in the local patterns most likely to defeat repair.

This separation is important. The paper does not claim a universal high-order guarantee. It claims a deterministic repair theorem conditional on component-graph connectivity and then measures how often re-rooting makes this condition true in high-order regimes. The heuristic has no approximation-ratio guarantee: an adversary can choose the four-neighbor or four-incident-link isolating cuts of Proposition~2. The experimental high-order claim is therefore distributional over the tested random, critical, near-source, and close-cluster placements. The close-cluster rows in Table~\ref{tab:rootcause} are the stress tests that most closely approximate adversarial local cuts.

\section{NoC-Oriented Evaluation Methodology}
\label{sec:evaluation}
The primary large-scale evaluation is graph-level and cycle-count oriented. This is intentional for the full $k\in\{10,25,50,100,200\}$ campaign: the goal is to test the mathematical recovery boundary, component connectivity, repair-edge count, candidate overhead, and transient discovery behavior over hundreds of thousands of trials. To strengthen the NoC interpretation, we also include patched native Gaussian-link Noxim scheduled replays. The main replay uses $k=25$ ($N=1301$) and an additional $k=50$ audit checks the larger-router case. In the patched build, the Noxim topology generator instantiates Gaussian physical links $u\leftrightarrow u\pm k$ and $u\leftrightarrow u\pm(k+1)\pmod N$; the routing module uses the same Gaussian coordinate-distance rule as the graph model. Thus the graph-level campaign is the large-scale evidence, while the Noxim replay is a focused router-level validation of recovered broadcast operation. It is not a standard unmodified Noxim topology and not a random saturation-throughput experiment; it is a scheduled replay of recovered broadcast and repair traces on a documented Gaussian-link Noxim patch.

\subsection{Experimental Settings}
We test
\begin{equation}
k\in\{10,25,50,100,200\},
\end{equation}
corresponding to $N=221,1301,5101,20201,80401$ nodes. Static experiments use 280,000 trials. Transient experiments use 15,000 trials. Static fault regimes include node-only, link-only, and mixed node/link faults under random, critical, near-source, and close placements. Transient trials include critical-near-root, random-tree-edge, and boundary-branch failed links.

\subsection{Compared Recovery Modes}
Four modes are evaluated.
\begin{enumerate}
\item \emph{Baseline no repair:} the original source broadcasts on the original tree; failed forwarding nodes or links stop propagation.
\item \emph{Re-root avoid-only:} select a new source and succeed only if its tree avoids failed links and places node faults as leaves.
\item \emph{Fixed-source component repair:} keep the original source, prune its tree, and repair components if the component graph is connected.
\item \emph{Re-root plus component repair:} select a damage-reducing root, prune the selected tree, and repair components. This is the proposed hybrid method.
\end{enumerate}

\subsection{Measured Metrics}
We measure success rate, reached healthy nodes, candidate count, candidates checked, scan exhaustion, relocation success, components after pruning, repair edges, repaired depth, total recovery steps, failed tree links used after re-rooting, and root-cause labels for failed hybrid trials. For NoC interpretation, the number of repair edges is also the number of exceptional component-crossing forwarding entries required by the repaired tree. Thus reducing repair edges reduces additional forwarding state and exceptional control actions. The native Noxim replay additionally reports completion cycles, control packets, forwarding exceptions, router delay, throughput, and within-simulator energy per packet.

\subsection{Patched Native Gaussian-Link Noxim Scheduled Replay}
The native Noxim validation uses the same $k=25$ network ($N=1301$), but it does not rely on a standard Noxim mesh, torus, or butterfly topology. We used a patched Noxim build in which the topology generator was extended with a \texttt{GAUSSIAN} mode and the routing module was extended with a \texttt{GAUSSIAN} mode. For $k=25$, the simulator instantiates 1301 routers indexed by $u\in\mathbb{Z}_{1301}$ and creates the physical adjacency
\begin{equation}
\mathrm{Adj}(u)=\{u\pm 25,\;u\pm 26\}\pmod {1301} .
\end{equation}
The routing mode forwards over these native Gaussian links using the same canonical-coordinate distance rule used in the graph model. The command line therefore uses \texttt{-topology GAUSSIAN}, \texttt{-dimx 1301}, \texttt{-dimy 1}, \texttt{-routing GAUSSIAN}, and a hardcoded scheduled-replay traffic file. These flags are not part of unmodified Noxim; they refer to the documented patch used for this experiment.

Unlike the earlier table-probability workload, the scheduled replay is not a random saturation-throughput experiment. Each replay file injects recovery/control traffic and data packets according to the selected repaired broadcast structure. Thus the reported completion time includes recovery setup and scheduled broadcast completion. The purpose is to validate the broadcast recovery operation under router-level execution, not to characterize the saturation throughput of arbitrary application traffic.

Table~\ref{tab:noxim_patch} summarizes the Noxim patch and replay settings. The native replay is a focused collective-operation validation rather than a broad NoC scaling sweep. The $k=25$ run evaluates five regimes: fault-free, one critical tree link, five critical tree links, a higher-order clustered node-fault stress case, and a two-node/two-link mixed case. The compared methods are baseline no repair, fixed-source component repair, and hybrid re-rooting plus component repair. We run 30 trials per regime and method, for 450 Noxim executions. All native-link runs completed with zero nonzero return codes, zero parse warnings, zero packet mismatches, and zero scheduled-replay violations in the packet/replay audit. A $k=50$ audit also completed 450/450 Noxim runs, and the only candidate-window warning disappeared after full-root scanning. The output logs verify the use of \texttt{-topology GAUSSIAN}, \texttt{-dimx N}, \texttt{-dimy 1}, \texttt{-routing GAUSSIAN}, and hardcoded scheduled replay traffic.

\begin{table}[H]
\centering
\caption{Patched Noxim reproducibility summary for the native Gaussian-link replay.}
\label{tab:noxim_patch}
\footnotesize
\begin{tabular}{ll}
\toprule
Item & Value \\
\midrule
Simulator & Patched Noxim build \\
Topology flag & \texttt{-topology GAUSSIAN} \\
Routing flag & \texttt{-routing GAUSSIAN} \\
Routers & $N=1301$ main replay; $N=5101$ audit replay \\
Physical links & $u\leftrightarrow u\pm k$, $u\leftrightarrow u\pm(k+1)$ (mod $N$) \\
Traffic mode & Hardcoded scheduled broadcast/recovery replay \\
Regimes/methods/trials & $5\times3\times30=450$ runs per audit set \\
Parser status & 0 nonzero returns, 0 parse warnings, 0 packet mismatches \\
Audit status & packet/replay clean; $k=50$ full-root scan removes window artifact \\
Supplementary material & Patch specification, replay traces, logs, and audit tables \\
\bottomrule
\end{tabular}
\end{table}

\section{Results}
\label{sec:results}
\subsection{Claim-Regime Recovery}
\begin{figure}[H]
\centering
\includegraphics[width=0.7\linewidth]{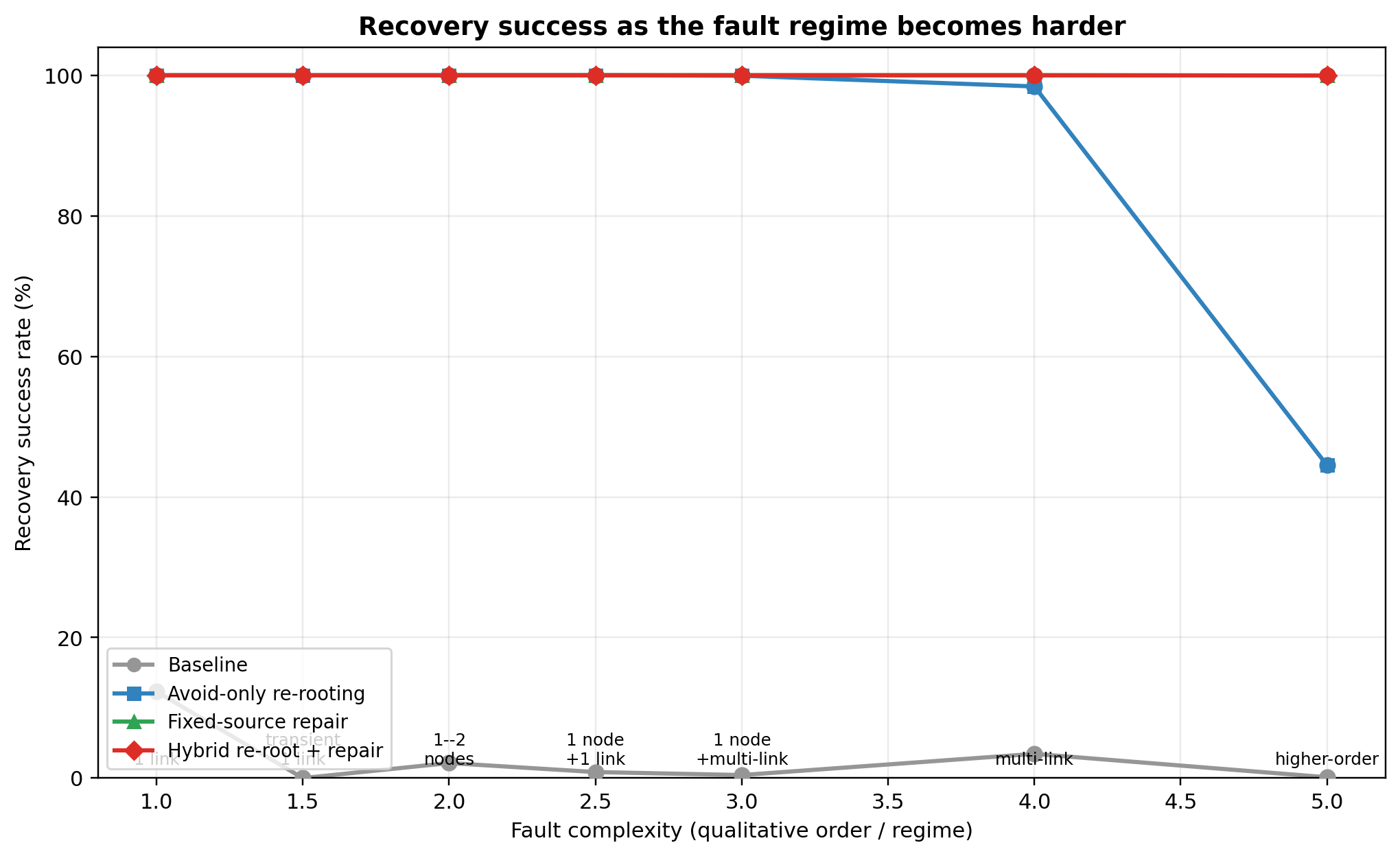}
\caption{Recovery success as the fault regime becomes harder. The figure summarizes the degradation curves for the baseline, avoid-only re-rooting, fixed-source component repair, and the proposed hybrid method. The hybrid curve remains essentially flat near 100\%, while avoid-only recovery degrades sharply in the higher-order regimes.}
\label{fig:successcurve}
\end{figure}

Table~\ref{tab:claim} summarizes the main recovery regimes. The first three rows are deterministic or theorem-candidate bounded regimes. The multi-link and high-order rows are empirical/best-effort regimes. Figure~\ref{fig:successcurve} complements the table by showing the four recovery curves as fault complexity increases.

\begin{table}[H]
\centering
\caption{Full-run recovery by claim regime.}
\label{tab:claim}
\begin{tabular}{lrrr}
\toprule
Regime & Trials & Avoid-only & Hybrid repair \\
\midrule
One/two node faults & 40,000 & 100.000\% & 100.000\% \\
Single failed link & 20,000 & 100.000\% & 100.000\% \\
One node + one link & 20,000 & 100.000\% & 100.000\% \\
One node + multiple links & 20,000 & 99.935\% & 100.000\% \\
Multiple links & 60,000 & 98.408\% & 99.998\% \\
Higher-order heuristic & 120,000 & 44.498\% & 99.963\% \\
Transient single link & 15,000 & 100.000\% & 100.000\% \\
\bottomrule
\end{tabular}
\end{table}

\subsection{Repair-Edge Reduction}
Table~\ref{tab:repair_edges} gives the strongest empirical evidence. Fixed-source component repair usually succeeds, but it uses substantially more repair edges. Re-rooting reduces average repair edges to zero in the deterministic node and single-link regimes and remains much lower in mixed and high-order regimes.

\begin{table}[H]
\centering
\caption{Average external repair-edge reduction and measured hybrid repaired depth, aggregated over all tested $k$. The last column reports the measured average depth of the re-rooted hybrid repaired tree, expressed relative to $k$; it is not inferred from $R$ and is not a bound for the fixed-source repaired tree.}
\label{tab:repair_edges}
\begin{adjustbox}{max width=\textwidth}
\begin{tabular}{lrrrr}
\toprule
Scenario & Fixed-source edges & Re-root edges & Reduction & Hybrid avg. depth \\
\midrule
1 node & 1.063 & 0.000 & 100.00\% & $k$ exact \\
2 nodes & 2.098 & 0.000 & 100.00\% & $k$ \\
1 link & 0.877 & 0.000 & 100.00\% & $k$ \\
2 links & 1.751 & 0.000 & 100.00\% & $k$ \\
3 links & 2.617 & 0.00025 & 99.99\% & $k+0.00025$ \\
5 links & 4.374 & 0.047 & 98.92\% & $k+0.047$ \\
1 node + 1 link & 1.937 & 0.000 & 100.00\% & $k$ \\
1 node + 2 links & 2.800 & 0.00065 & 99.98\% & $k+0.00065$ \\
2 nodes + 1 link & 2.966 & 0.197 & 93.37\% & $k+0.197$ \\
2 nodes + 2 links & 3.838 & 0.477 & 87.58\% & $k+0.477$ \\
3 nodes + 2 links & 4.841 & 0.971 & 79.94\% & $k+0.971$ \\
\bottomrule
\end{tabular}
\end{adjustbox}
\end{table}

The last column of Table~\ref{tab:repair_edges} reports the measured average depth of the selected hybrid repaired tree, not a quantity computed from the repair-edge count $R$. This notation is used only to summarize how close the hybrid depth remains to the fault-free diameter $k$ in the tested regimes. Lemma~4 bounds the number of components and hence the minimum number of exceptional component-crossing rules; the longest hop path also depends on where those crossing edges attach inside the affected components. This distinction is essential in the fixed-source rows of Table~\ref{tab:noxim_native}, where a small number of repair edges can reconnect deep subtrees and therefore produce depths much larger than $k+R$. A latency-dominant implementation using $J_{\alpha,\beta,\gamma}$ can therefore read the edge and depth measurements together: when $\beta\gg\alpha$, a different root weighting may be preferable, and the tables provide the data needed to evaluate that tradeoff without rerunning the reachability experiments.

\begin{table}[H]
\centering
\caption{Analytical component-count/repair bound compared with observed averages.}
\label{tab:component_bound_eval}
\begin{tabular}{lrrr}
\toprule
Fault class & Bound & Observed avg. & Ratio \\
\midrule
1 node & 4 & 1.063 fixed / 1.000 rerooted & 0.27 / 0.25 \\
2 nodes & 7 & 2.098 fixed / 1.000 rerooted & 0.30 / 0.14 \\
3 links & 4 & 2.617 fixed / 0.00025 rerooted & 0.66 / 0.00006 \\
5 links & 6 & 4.374 fixed / 0.047 rerooted & 0.73 / 0.008 \\
\bottomrule
\end{tabular}
\end{table}

The bound in Lemma~4 is tight in the worst case but loose in practice. It is tightest for link faults because each failed tree link can create exactly one additional component; it is loosest for re-rooted node faults because leaf placement suppresses most component creation.

\subsection{Single-Link Theorem-Mining Results}
For all tested $k$ and placements, the single-link regime has 100\% candidate found, 100\% selected-source link safety, and 100\% hybrid recovery. This matches Theorem~3. The result is not merely empirical: single-edge deletion cannot disconnect $\Gk$, and if the failed edge lies in the selected tree, it creates exactly two tree components that are reconnectable by one crossing edge.

\subsection{Transient Recovery}
Table~\ref{tab:transient} reports transient one-link recovery. Discovery layer varies by placement, showing that the simulation records actual detection depth rather than always charging the full diameter. Hybrid recovery succeeds in every transient trial. The discovery-layer values supply the timing inputs for the recovery-latency decomposition.

\begin{table}[H]
\centering
\caption{Transient one-link recovery summary.}
\label{tab:transient}
\begin{tabular}{lrrr}
\toprule
Placement & Trials & Success & Avg. discovery layer \\
\midrule
Critical near root & 5,000 & 100.000\% & 14.10 \\
Random tree edge & 5,000 & 100.000\% & 51.42 \\
Boundary branch & 5,000 & 100.000\% & 76.52 \\
\bottomrule
\end{tabular}
\end{table}

\subsection{Root-Cause Analysis of Non-Perfect Rows}
The two non-100\% rows in Table~\ref{tab:claim} are fully explained by the component-graph condition. Table~\ref{tab:rootcause} breaks down all hybrid failures. In the multi-link regime, there is one failure out of 60,000 trials: a $k=10$ five-link near-source case with a disconnected healthy component graph. In the higher-order heuristic regime, there are 44 failures out of 120,000 trials. Forty-three have successful relocation but disconnected component graphs; one has both a disconnected component graph and failed relocation. The failures are concentrated in five-node close or near-source clustered cuts at small $k$. These are exactly the patterns expected to be difficult: several faults remove neighboring forwarding and bypass opportunities at once, so the healthy graph may remain large but the component graph induced by the selected tree can become disconnected.

\begin{table}[H]
\centering
\caption{Root-cause breakdown of all non-recovered static hybrid trials.}
\label{tab:rootcause}
\footnotesize
\begin{adjustbox}{max width=\textwidth}
\begin{tabular}{p{0.48\linewidth}rrrr}
\toprule
Regime/case & Trials & Failures & Disconnected $\Cgraph_r$ & Relocation fail \\
\midrule
Multiple links, all & 60,000 & 1 & 1 & 0 \\
\quad $k=10$, 5 links, near-source & 4,000 & 1 & 1 & 0 \\
Higher-order heuristic, all & 120,000 & 44 & 44 & 1 \\
\quad 5 nodes, close, $k=10$ & 1,000 & 24 & 24 & 0 \\
\quad 5 nodes, close, $k=25$ & 1,000 & 7 & 7 & 0 \\
\quad 5 nodes, close, $k=50$ & 1,000 & 4 & 4 & 0 \\
\quad 5 nodes, close, $k=100$ & 1,000 & 1 & 1 & 0 \\
\quad 5 nodes, close, $k=200$ & 1,000 & 4 & 4 & 0 \\
\quad 5 nodes, near-source, $k=10$ & 1,000 & 3 & 3 & 1 \\
\quad 2 nodes + 3 links, near-source, $k=10$ & 1,000 & 1 & 1 & 0 \\
\bottomrule
\end{tabular}
\end{adjustbox}
\end{table}

This table is important for interpretation. The non-100\% rows are not unexplained anomalies. They occur precisely where the theorem allows failure: the healthy component graph is disconnected or the relocation path is unavailable. Thus the high-order rows should be interpreted as empirical best-effort recovery rather than deterministic guarantees.

The concentration in close five-node clusters has a structural explanation. A close cluster removes several adjacent side-entry and bypass ports at once. In small $k$, the boundary is near every local cut, so there are fewer alternate component-crossing edges after quotient identification. Proposition~2 gives the limiting obstruction: four neighboring node faults can isolate a healthy vertex, and four incident link faults can do the same. The observed five-node close failures are not exactly that trivial obstruction in every trial, but they are the same phenomenon at the component level: all crossing edges from one pruned component to the rest of the healthy graph are removed.

\subsection{Candidate-Generation Overhead}
Table~\ref{tab:candidate_complexity} stated the algorithmic bound. In the full run, the exact one/two node source selection checked one candidate in the guaranteed node rows because the boundary intersection returned a valid candidate immediately. Proposition~3 reduces each failed-link test to constant-time coordinate comparisons, so the link-only candidate phase checks roots without constructing full trees. Link-only avoid-only scans sometimes reached the 20,000 cap for $k=100$ and $k=200$ multi-link trials, but the proposed hybrid repair still achieved 100\% recovery in those large-$k$ multi-link rows because component repair does not require perfect link avoidance. Therefore scan-cap effects influence the avoid-only comparator, not the correctness of the component-repair phase. Table~\ref{tab:link_candidate_overhead} reports the actual link-only candidate overhead. The average remains far below the cap for one to three failed links. Five-link cases are the only setting where the cap is sometimes reached; this is exactly the regime already labeled empirical rather than guaranteed.

\begin{table}[H]
\centering
\caption{Candidate checks for link-only avoid-only selection in the full run.}
\label{tab:link_candidate_overhead}
\begin{tabular}{rrrrr}
\toprule
Failed links & Trials & Avg. checked & 95th pct. & Cap hits \\
\midrule
1 & 20,000 & 9.20 & 41.0 & 0 \\
2 & 20,000 & 16.75 & 48.0 & 0 \\
3 & 20,000 & 95.72 & 80.0 & 0 \\
5 & 20,000 & 681.34 & 4527.25 & 320 \\
\bottomrule
\end{tabular}
\end{table}

\subsection{Patched Native Gaussian-Link Noxim Scheduled Replay}
Table~\ref{tab:noxim_native} reports the patched native Gaussian-link Noxim scheduled replay. The $k=25$ run contains 450 Noxim executions: five regimes, three methods, and 30 trials per regime/method. All runs completed successfully; the audit reported zero nonzero return codes, zero parse warnings, zero packet mismatches, zero failed-link uses in the scheduled trace, zero faulty endpoints, zero parent-order violations, and zero non-Gaussian replay edges. The link-only sanity checks confirm that the hybrid selector changes the source in all one-link and five-link critical trials. A follow-up $k=50$ audit produced the same packet/replay/parser outcome; its only warning was traced to scanning roots $0,\ldots,4999$ in a network with $N=5101$, and a full-root scan selected root 2475 with zero failed tree links, zero repair edges, complete reach $5101/5101$, and Noxim return code 0.

\begin{table}[H]
\centering
\caption{Patched native Gaussian-link Noxim scheduled broadcast replay for $k=25$ ($N=1301$), 30 trials per regime/method. The patched topology and routing use physical links $u\leftrightarrow u\pm k$ and $u\leftrightarrow u\pm(k+1)$ modulo $N$. Dynamic energy per packet is reported in units of $10^{-11}$ J and is interpreted only as a within-simulator comparison under a common power file. Completion cycles include recovery setup and scheduled broadcast completion.}
\label{tab:noxim_native}
\footnotesize
\begin{adjustbox}{max width=\textwidth}
\begin{tabular}{llrrrrrr}
\toprule
Regime & Method & Success & Depth & Repair edges & Completion & Avg. delay & Throughput \\
\midrule
Fault-free & Baseline & 100.00\% & 25.00 & 0.00 & 40.00 & 4.3200 & 2.514510 \\
One critical link & Fixed-source & 100.00\% & 42.87 & 1.00 & 58.33 & 4.3198 & 2.476280 \\
One critical link & Hybrid & 100.00\% & 25.00 & 0.00 & 58.60 & 4.3405 & 2.493610 \\
Five critical links & Fixed-source & 100.00\% & 63.10 & 5.00 & 81.90 & 4.3235 & 2.430110 \\
Five critical links & Hybrid & 100.00\% & 25.13 & 0.03 & 67.10 & 4.3484 & 2.483250 \\
Higher-order & Fixed-source & 100.00\% & 33.43 & 3.20 & 51.63 & 4.3248 & 2.490480 \\
Higher-order & Hybrid & 100.00\% & 25.00 & 0.00 & 61.60 & 4.3459 & 2.488260 \\
Two nodes + two links & Fixed-source & 100.00\% & 51.63 & 4.37 & 70.20 & 4.3257 & 2.453100 \\
Two nodes + two links & Hybrid & 100.00\% & 25.13 & 0.20 & 74.20 & 4.3549 & 2.472640 \\
\bottomrule
\end{tabular}
\end{adjustbox}

\vspace{0.5ex}
\begin{adjustbox}{max width=\textwidth}
\begin{tabular}{llrrrr}
\toprule
Regime & Method & Components & Control packets & Network tail & Dyn. energy/packet \\
\midrule
One critical link & Fixed-source & 2.00 & 1.00 & 15.47 & 1.2424 \\
One critical link & Hybrid & 1.00 & 1.00 & 33.60 & 1.2489 \\
Five critical links & Fixed-source & 6.00 & 5.00 & 14.80 & 1.2424 \\
Five critical links & Hybrid & 1.03 & 1.03 & 41.93 & 1.2514 \\
Higher-order & Fixed-source & 4.20 & 3.20 & 16.00 & 1.2424 \\
Higher-order & Hybrid & 1.00 & 1.00 & 36.60 & 1.2503 \\
Two nodes + two links & Fixed-source & 5.37 & 4.37 & 15.20 & 1.2424 \\
Two nodes + two links & Hybrid & 1.20 & 1.20 & 48.87 & 1.2533 \\
\bottomrule
\end{tabular}
\end{adjustbox}
\end{table}

The native replay confirms the structural graph-level trend under router-level execution while showing why completion cycles should not be overclaimed. Hybrid reduces repair edges from 1.00 to 0.00, 5.00 to 0.03, 3.20 to 0.00, and 4.37 to 0.20 in the four faulty regimes, and it keeps repaired depth near $k$, consistent with Proposition~5. Completion cycles depend on setup and delivery tail: boundary re-rooting can reduce components while spreading the last deliveries across a wider coordinate span than a fixed central source. Hybrid improves the five-link critical completion time by about 18\%, is essentially tied in the one-link case, and is slower in the higher-order and two-node/two-link rows under the state objective. Table~\ref{tab:noxim_audit} confirms that zero setup makes the hybrid selectors faster or competitive in all audited regimes, while the latency-weighted selector improves normal completion in the two non-dominant regimes. Thus the safe conclusion is 100\% delivery with far fewer repair edges, components, and depth; completion-cycle benefit is regime- and objective-dependent rather than universal.

\begin{table}[H]
\centering
\caption{$k=25$ completion-cycle audit separating setup and selector objective. The zero-setup rows now include the critical one-link and five-link regimes; all rows had zero packet mismatches, parse warnings, return-code failures, failed-link replay uses, and replay-tree violations.}
\label{tab:noxim_audit}
\begin{tabular}{llrrr}
\toprule
Regime & Setup & Fixed & Hybrid-state & Hybrid-latency \\
\midrule
1 critical link & zero & 56.0 & 40.0 & 40.0 \\
5 critical links & zero & 79.0 & 40.0 & 40.0 \\
Higher-order & normal & 49.6 & 58.2 & 54.8 \\
Higher-order & zero & 46.8 & 40.0 & 40.0 \\
2 nodes + 2 links & normal & 75.6 & 73.8 & 71.8 \\
2 nodes + 2 links & zero & 71.8 & 40.2 & 40.0 \\
\bottomrule
\end{tabular}

\footnotesize Critical-regime normal rows appear in Table~\ref{tab:noxim_native}; zero setup isolates reconfiguration overhead.
\end{table}

\section{Discussion}
\label{sec:discussion}
\subsection{Why the Contribution Is Not Merely Incremental}
The hybrid method changes the objective from either source selection or local tree repair alone to damage-minimizing root selection followed by optimal residual repair. Static re-rooting asks whether a root exists that places node faults on the boundary. Component repair asks how to reconnect a given pruned tree. The present paper asks a different question: which root should be selected so that the post-fault component structure is as small as possible, and how should residual link and transient failures be repaired with minimum external edges? This is a new coupled optimization because the chosen root changes the component graph on which the $c-1$ theorem acts. The empirical consequence is visible in Table~\ref{tab:repair_edges}: fixed-source repair often succeeds, but re-rooting changes the component partition and sharply reduces required repair edges. This is why the primary empirical metric is repair-edge reduction, not only success rate.

\subsection{NoC Interpretation}
In a NoC implementation, a repair edge corresponds to one exceptional forwarding action between healthy components; hence Table~\ref{tab:repair_edges} measures additional forwarding-state cost. The patched native Gaussian-link Noxim replay adds router-level timing, throughput, and within-simulator energy evidence, while the main claim remains recovery correctness and reduction of exceptional repair state. Tables~\ref{tab:noxim_native} and~\ref{tab:noxim_audit} show that this advantage can reduce scheduled completion cycles, especially with zero setup or a latency-weighted selector, but completion time is not universally minimized by the state-sensitive selector. The longer tail has a geometric source: distance-$k$ boundary placement can make the final delivery frontier span the full coordinate-ball diameter even when the component count is minimal. Calibrated EDP, saturation throughput, background traffic, and virtual-channel pressure are natural extensions, not assumptions needed for the current edge-minimum repair claim.

\subsection{Why the Graph Theorems Matter}
The component-repair theorem is elementary in its final graph form, but its role in the paper is structural: it converts every complicated node/link fault pattern into a component graph whose repair optimum is known exactly. The nontrivial design question is therefore moved to root selection and damage reduction. This separation is useful for TC-style interconnection work because it prevents empirical success from hiding nonminimal repair. Every successful hybrid trial is accompanied by a certificate: a selected root, a component graph, and exactly $c-1$ crossing edges.

\subsection{Limits}
The deterministic guarantees are intentionally limited. One/two node re-rooting and one-link repair are guaranteed. For multiple links and high-order node faults, the theorem is conditional on the connectedness of $\Cgraph_r$. The root-cause table shows exactly where this condition fails. The failure concentration at small $k$ has a quantitative explanation. At $k=10$, the network has $N=221$ nodes; five close-cluster faults can remove five nodes and many adjacent forwarding opportunities, representing a visible fraction of the local bypass capacity. At $k=200$, the same five faults represent only $5/80{,}401\approx0.006\%$ of the network, so the relative bypass-capacity loss is negligible and the component graph almost always remains connected. For fixed fault count, this density scales as $O(1/k^2)$ because $N=2k^2+2k+1$. This explains why the failures in Table~\ref{tab:rootcause} are concentrated in small-$k$ close clusters. Proposition~4 gives a first-order union-bound explanation for fixed-size local obstructions, but a complete high-probability connectivity theorem proving that $\Cgraph_r$ is connected with high probability under random uniform faults, for example for $q=o(k)$, remains open. The transient evaluation is limited to one link discovered during propagation; node transients, sequential multi-link discoveries, and cascading recovery-phase faults are outside scope. A complete characterization of high-order cuts that defeat repair, and a constant-size link-safety selector analogous to the 144-case node selector, remain open.

\subsection{Future Cycle-Accurate Evaluation}
The patched native Gaussian-link Noxim replay removes the main topology limitation of the earlier workload-only validation, and the $k=25/k=50$ audits close the packet-accounting and replay-validity concerns. Larger $k$ sweeps, more injection rates, background traffic, virtual-channel pressure, calibrated power models, and EDP remain future NoC extensions rather than prerequisites for the present scheduled collective-operation validation.

\section{Complexity}
For a fixed root, constructing the tree, pruning it, computing connected components, scanning Gaussian neighbor edges, and building a component spanning tree all take $O(N)$ time because the network has degree four. Candidate generation costs $O(k)$ for exact one/two node boundary enumeration in the validation prototype, $O(1)$ for the shifted sign-case primitive, and $O(M|\Fe|)$ for link-only candidate testing with $M\le20000$ in the full run. Therefore the full experimental implementation is $O(N+M|\Fe|)$ per trial for a selected root, not $O(N^2)$. Table~\ref{tab:link_candidate_overhead} shows that the average link-only $M$ is much smaller than the cap except in five-link stress cases. Lemma~4 bounds the component count and hence the number of external repair rules; Proposition~5 gives the corresponding coarse depth accounting, while the exact repaired depth is measured separately because it also depends on the attachment endpoints and the final rooted orientation.

In a hardware implementation using the constant-size two-node generator and a bounded number of link faults, the candidate phase is constant size and the repair phase can be restricted to the affected components or subtree. The global implementation used here is a conservative validation model.

\section{Conclusion}
\label{sec:conclusion}
This paper introduced re-rooting-assisted edge-minimum runtime repair for dense Gaussian broadcast networks under static node faults, static link faults, mixed static faults, and runtime-discovered single-link faults. The method selects a damage-reducing source and then repairs residual fragmentation. If the healthy component graph has $c$ components and is connected at the component level, exactly $c-1$ external repair edges are necessary and sufficient; a single failed Gaussian link is always avoided or repaired with at most one crossing edge.

Experiments show that the hybrid method recovers all tested deterministic and bounded mixed regimes, recovers 99.998\% of multi-link trials and 99.963\% of higher-order heuristic trials, and reduces repair-edge count by 80--100\% compared with fixed-source repair. The few failures are fully explained by disconnected component graphs or relocation failure. Patched native Gaussian-link Noxim scheduled replay and audit runs show that this structural reduction yields smaller components and shallower repaired structures under router-level execution. The completion-cycle audit confirms that the replay and parser are clean while also showing that completion time depends on setup, scheduling, delivery tail, and selector objective. Thus completion time is reported as an objective-dependent metric rather than a universal dominance claim.

\section*{Data and Reproducibility}
The run uses fixed seeds and writes raw trials, summaries, metadata, and figure-ready CSV outputs. A supplementary artifact package contains graph scripts, raw CSVs, plotting inputs, the patched-Noxim driver, native Gaussian-link patch, scheduled traces, logs, parsers, packet/replay audit tables, and the full-root-scan $k=50$ correction record, prepared for DOI-linked archival deposition.

\end{document}